\IfFileExists{Definitions/mdpi.cls}{%
  \documentclass[particles,review,submit,moreauthors]{Definitions/mdpi}%
}{%
  \documentclass[particles,review,submit,moreauthors]{mdpi_fallback}%
}

\usepackage{mathtools,bm}
\graphicspath{{figures/}{Definitions/}}
\newcommand{\cevns}{\ensuremath{\mathrm{CE}\nu\mathrm{NS}}}
\newcommand{\Lagr}{\mathcal{L}}

\AtBeginDocument{\setlength{\headheight}{20pt}}

\firstpage{1}

\pubvolume{1}
\issuenum{1}
\articlenumber{0}
\pubyear{2026}
\copyrightyear{2026}
\datereceived{}
\daterevised{}
\dateaccepted{}
\datepublished{}

\Title{Light Mediators at the Neutrino--Dark Matter Frontier: Coherent Scattering, Precision Constraints, and Complementary Searches}

\Author{Raoul Serao $^{1,2,*}$, Mohamed Ouchemhou $^{3}$, Antonio Capolupo$^4$ and Salvatore Capozziello $^{1,2,5}$}

\AuthorNames{Raoul Serao, Mohamed Ouchemhou, Antonio Capolupo, Salvatore Capozziello}

\address{%
$^{1}$ Scuola Superiore Meridionale,
Largo San Marcellino 10, 80138 Napoli, Italy;
r.serao@ssmeridionale.it\\
$^{2}$ Istituto Nazionale di Fisica Nucleare,
Sezione di Napoli, Complesso Universitario
di Monte S. Angelo, Via Cintia Edificio 6,
80126 Napoli, Italy\\
$^{3}$ Ruđer Bošković Institute,
Bijenička cesta 54, 10000 Zagreb, Croatia;
ouchemhou2@gmail.com\\
$^4$Dipartimento di Fisica E.~R. Caianiello, Universit\`a di Salerno, and INFN -- Gruppo Collegato di Salerno,
Via Giovanni Paolo II 132, 84084 Fisciano (SA), Italy\\
$^{5}$ Dipartimento di Fisica ``Ettore Pancini'',
Complesso Universitario di Monte S. Angelo,
Università degli Studi di Napoli ``Federico II'',
Via Cintia Edificio 6, 80126 Napoli, Italy;
capozziello@na.infn.it\\

}

\corres{Correspondence: r.serao@ssmeridionale.it}

\abstract{Low-threshold neutrino and dark-matter experiments constrain light mediators through different momentum transfers, targets, and coupling combinations. We examine scalar and vector interactions in coherent neutrino--nucleus scattering, neutrino--electron scattering, and dark-matter direct detection. Three representative interaction models are used to compare recoil spectra and coupling degeneracies. For $^{73}\mathrm{Ge}$ and $^{131}\mathrm{Xe}$, the nuclear-charge response matrix has a condition number of approximately 115, so two different targets may still leave nearly parallel blind directions. A covariance-weighted extension separates charge geometry from measurement precision. Flux-integrated vector-mediator predictions and a physically derived solar-neutrino deformation quantify recoil-rate changes and local Fisher uncertainty inflation in the neutrino fog. We discuss the implications of measurements from COHERENT, CONUS+, LZ, XENONnT, and PandaX-4T and specify which theoretical inputs, detector responses, and statistical correlations are needed to combine their constraints.}

\begin{document}

\section{Introduction}

Low-energy scattering experiments provide a way to search for new interactions that may be difficult to isolate at higher energies. A mediator with a mass comparable to, or smaller than, the momentum transferred in a collision cannot in general be represented by a momentum-independent contact interaction. Its propagator changes the recoil-energy distribution, and the observable effect depends on the particles to which it couples. Neutrino scattering is particularly useful in this setting because neutrino sources cover different energies and flavors, while nuclear and electron targets probe different fermion currents. The same measurements are also relevant to dark-matter searches, where neutrino-induced recoils have become an experimentally observed background rather than only a projected one \cite{Dent2017,Coloma2021,Pandey2024}.

Coherent elastic neutrino--nucleus scattering (\cevns) is central to these studies. In the Standard Model (SM), the process proceeds through neutral-current exchange, with a leading coherent amplitude governed by the nuclear weak charge and, for many nuclei, dominated by its neutron contribution. The enhancement of the rate comes at the cost of a small nuclear recoil energy, making the detector threshold, nuclear response, and energy calibration essential parts of the measurement. Predicted by Freedman \cite{Freedman1974}, \cevns\ was first observed by COHERENT with a stopped-pion neutrino source \cite{Akimov2017}. Measurements on argon and CsI then provided information from targets with different proton-to-neutron ratios \cite{Akimov2021Argon,Akimov2022CsI}. The COHERENT germanium results reported in 2025 and 2026 further extended the available recoil and timing information \cite{COHERENT2025Ge,COHERENT2026Ge}. At lower neutrino energies, CONUS+ observed reactor-antineutrino \cevns\ in 2025 \cite{Ackermann2025,DeRomeri2025}. These measurements are not simply repetitions of the same cross-section test: stopped-pion and reactor sources differ in flavor composition, energy distribution, and accessible momentum transfer. Their ability to distinguish new interactions depends on those differences as well as on the precision of each dataset.

The connection with dark-matter detection has become more direct. PandaX-4T and XENONnT reported indications of solar $^8$B neutrinos through nuclear recoils in 2024 \cite{Bo2024PandaX,Aprile2024XENON}. In 2026, LZ and XENONnT reported solar-neutrino \cevns\ signals at $4.5\sigma$ and $3.3\sigma$, respectively \cite{Akerib2026LZ,Aprile2026XENON}. XENONnT also measured solar-neutrino--electron scattering at keV-scale recoil energies, with a signal dominated by $pp$ neutrinos \cite{Aprile2026Electron}. Liquid-xenon detectors therefore sample both nuclear and electron recoil channels, although the underlying neutrino spectra and detector responses are different. Solar-neutrino nuclear recoils overlap the region in which light dark matter is sought. This overlap gives rise to the so-called neutrino fog: a loss of dark-matter discovery sensitivity whose severity depends on the similarity of the recoil spectra and on uncertainties in the neutrino contribution \cite{Billard2014,OHare2021}. It is not a universal exposure-independent floor, and its interpretation changes if neutrino interactions depart from the SM.

Light scalar and vector mediators affect these observations in different ways. For a vector coupled to neutrinos and quarks, coherent scattering probes a product of neutrino and nuclear couplings divided by a momentum-dependent propagator. Flavor-diagonal amplitudes may interfere constructively or destructively with the SM contribution; the interference sign can change the allowed parameter regions and may produce cancellations in particular targets \cite{Coloma2020,Cadeddu2021,AtzoriCorona2026}. Chirality-changing scalar exchange, under the corresponding neutrino-state assumptions, instead contributes with a different recoil dependence and without leading interference with the SM vector amplitude. The contact-interaction description is appropriate only when the mediator mass is large compared with the relevant momentum transfer. For lighter states, the relative sensitivity of two experiments can change across recoil energy even at fixed couplings \cite{Dent2017,Coloma2021,Li2026EFT}. A comparison based only on mediator mass or on a quoted upper limit can therefore obscure the physical assumptions behind the result.

A related difficulty is that most recoil measurements do not determine individual couplings. Nuclear recoils constrain combinations of neutrino and quark couplings, with the up- and down-quark contributions weighted by the proton and neutron content of the target. Electron recoils instead involve neutrino--electron coupling products. Reactor antineutrinos primarily test the electron flavor, stopped-pion sources provide several identifiable flavor components, and solar measurements contain flavor mixtures shaped by oscillations. A second target or source can reduce a degeneracy only if its response differs sufficiently from that of the first. In particular, two nuclei with similar proton fractions may constrain nearly parallel combinations of up- and down-quark couplings even when their event samples are statistically independent. Flavor and target complementarity must thus be assessed in parameter space, rather than inferred from the number of available experiments \cite{Coloma2017,ColomaDenton2017,Coloma2020,Li2025Combined}.

Other observables can supply information that scattering alone cannot provide, but their interpretation requires care. Neutrino oscillations probe forward-scattering matter potentials, whereas recoil experiments involve finite momentum transfer; the relation between the two depends on the mediator mass and charge assignments \cite{Wolfenstein1978,FarzanTortola2018,Coloma2021}. Magnetic moments, atomic spectroscopy, and other precision measurements can restrict charged-lepton interactions in particular light-vector models \cite{Serao2025X17Impact,Serao2026g2}. Studies of X17-like states illustrate why bounds derived from lepton observables cannot automatically be applied to neutrino--quark scattering without specifying how the relevant couplings are related. Conversely, a mediator with an additional dark-sector coupling can affect dark-matter and neutrino recoil rates through different coupling products. Effective interaction models are useful for separating these dependencies, while gauge-invariant realizations may impose additional relations and constraints.

Existing reviews and phenomenological studies have examined \cevns, non-standard neutrino interactions, and light-mediator limits from individual or combined datasets \cite{Pandey2024,CEvNSWhitepaper,AtzoriCorona2026,Li2025Combined}. Effective-field-theory treatments organize the allowed interaction structures \cite{Li2026EFT}, and studies of the neutrino fog quantify the resulting limitations for dark-matter discovery \cite{Billard2014,OHare2021}. For comparing experimental results, however, the relevant question is not only how stringent an exclusion curve appears. It is which parameter combinations are identified by the measured spectra, which degeneracies remain, and which uncertainties are shared between experiments. Thresholds, quenching or ionization yields, nuclear form factors, neutrino fluxes, and statistical prescriptions enter this assessment. Published constraints obtained after fixing different couplings or profiling different nuisance parameters need not describe the same region of a common model.

In this review, we examine scalar and vector interactions across \cevns, neutrino--electron scattering, and dark-matter direct detection. Three representative interaction models (V1, V2, and S1) are used to make their coupling assumptions and recoil signatures explicit. We compare the momentum-transfer ranges and the coupling products probed by current measurements, and use a two-target nuclear-charge calculation to quantify how an apparent complementarity can leave nearly parallel blind directions. We then study the effect of uncertain neutrino interactions on dark-matter sensitivity with an analytic profile-likelihood model that separates spectral overlap from external information on the neutrino background. The calculation is illustrative rather than a global fit to experimental data. Our purpose is to state precisely what can be inferred from each class of observable and what additional model, detector, and likelihood information is required before their constraints can be combined.

\section{Interaction Models and Scattering Rates}
\label{sec:model}

Coherent elastic neutrino--nucleus scattering (\cevns) provides a sensitive probe of weak interactions and light mediators. In the spin-independent coherent limit, the leading Standard Model (SM) cross section is
\begin{equation}
\label{eq:sm}
\frac{d\sigma_{\mathrm{SM}}}{dT_N}
\simeq
\frac{G_F^2 M}{4\pi}
Q_W^2
\left(1-\frac{MT_N}{2E_\nu^2}\right)
F_W^2(q^2),
\qquad q^2\simeq 2MT_N,
\end{equation}
where $E_\nu$ is the neutrino energy, $T_N$ the nuclear recoil energy, and $M$ the nuclear mass. We use the signed weak charge $Q_W\simeq(1-4\sin^2\theta_W)Z-N=2[Zg_V^p+Ng_V^n]$, with $g_V^p=\tfrac12-2\sin^2\theta_W$ and $g_V^n=-\tfrac12$. The frequently used neutron-positive weak charge is $-Q_W$; either sign gives the same SM cross section, but the relative mediator shift must follow the same convention. The form factor $F_W(q^2)$ describes the nuclear response. Equation~\eqref{eq:sm} neglects radiative and higher-order recoil corrections \cite{Freedman1974,Pandey2024,Li2026EFT}. Precision calculations require nuclear form factors and their uncertainties \cite{Helm1956,KleinNystrand1999,AbdelKhaleq2024}.

Coherence requires approximately $qR_N\lesssim 1$, where $R_N$ is the nuclear radius. The momentum transfer depends on both recoil energy and nuclear mass. Reactor experiments, stopped-pion sources, and solar-neutrino measurements therefore probe different momentum ranges. Experimental thresholds and efficiencies further modify the accessible distributions. Each measurement constrains an integrated recoil spectrum rather than a single momentum transfer.

Consider a neutral vector mediator $X_\mu$ with mass $m_X$ and interactions
\begin{equation}
\label{eq:vector}
\Lagr_X=
X_\mu\left[
\sum_{\alpha,\beta}
g_\nu^{\alpha\beta}
\bar\nu_\alpha\gamma^\mu P_L\nu_\beta
+g_e\bar e\gamma^\mu e
+g_u\bar u\gamma^\mu u
+g_d\bar d\gamma^\mu d
+g_\chi\bar\chi\gamma^\mu\chi
\right].
\end{equation}
The neutrino couplings satisfy
$g_\nu^{\beta\alpha}=(g_\nu^{\alpha\beta})^*$.
The charged-fermion couplings are taken real. The diagonal vector current of $\chi$ assumes Dirac dark matter; it vanishes for a single Majorana fermion. Equation~\eqref{eq:vector} defines an effective interaction model. Gauge invariance, anomaly cancellation, and ultraviolet (UV) completion can impose additional relations among its couplings \cite{Cadeddu2021,Coloma2021}. The X17 literature illustrates the role of charge assignments and kinetic mixing in specific light-vector models \cite{Serao2026X17Tensions,Serao2026X17Review}.

At zero momentum transfer, the mediator couples coherently to the nuclear charge
\begin{equation}
\label{eq:qx}
Q_X=Z(2g_u+g_d)+N(g_u+2g_d).
\end{equation}
At finite momentum transfer, the proton and neutron contributions acquire separate form factors:
\begin{equation}
\label{eq:qxform}
Q_X(q^2)=
Z(2g_u+g_d)F_p(q^2)
+N(g_u+2g_d)F_n(q^2).
\end{equation}
Here $F_p(0)=F_n(0)=1$. Differences between proton and neutron distributions affect precision constraints. Nuclear structure, isotope composition, and higher-order responses must also be considered \cite{Hoferichter2020,Li2026EFT,Anand2014}.

Assuming identical proton and neutron form factors, the flavor-diagonal effective weak charge becomes
\begin{equation}
\label{eq:shift}
Q_{\mathrm{eff}}^\alpha(q^2)
=
Q_W+
\frac{\sqrt{2}\,g_\nu^{\alpha\alpha}Q_X}
{G_F(q^2+m_X^2)}.
\end{equation}
The normalization and sign can be checked by integrating out $X_\mu$. With $j_\nu^\mu=\bar\nu\gamma^\mu P_L\nu$ and the coherent nuclear current $J_N^\mu$, the coefficients in the common-form-factor approximation are
\begin{equation*}
\mathcal L_{\nu N}^{\rm eff}=-\left[\frac{G_FQ_W}{\sqrt{2}}+
\frac{g_\nu^{\alpha\alpha}Q_X}{q^2+m_X^2}\right]j_\nu^\mu J_{N\mu}.
\tag{5a}\label{eq:vector_matching}
\end{equation*}
Factoring $-G_F/\sqrt{2}$ gives Eq.~\eqref{eq:shift}, including its factor of $\sqrt{2}$ for the $P_L$ neutrino current. Keeping a neutron-positive $Q_W$ would instead require a minus sign before the mediator term. This is consistent with the signed vector-charge convention used in Ref.~\cite{AtzoriCorona2022Light}. Flavor-diagonal vector interactions can interfere with the SM amplitude. Flavor-changing interactions instead contribute incoherently when the final neutrino flavor differs from the initial one. Destructive interference can suppress the recoil rate and generate disconnected allowed regions.

The corresponding momentum-dependent effective coupling is
\begin{equation}
\label{eq:epsilon}
\varepsilon_{\alpha\alpha}^{q,V}(q^2)
=
\frac{g_\nu^{\alpha\alpha}g_q}
{2\sqrt{2}G_F(q^2+m_X^2)}.
\end{equation}
Here the coefficient is defined by $\mathcal L_{\rm NSI}=-2\sqrt{2}G_F\varepsilon_{\alpha\alpha}^{q,V}j_{\nu,\alpha}^\mu(\bar q\gamma_\mu q)$. With the usual separate chiral-quark operators, it equals each of $\varepsilon^{q,L}=\varepsilon^{q,R}$; their vector sum is $\varepsilon^{q,L}+\varepsilon^{q,R}=2\varepsilon^{q,V}$ in this notation. This distinction must be retained when translating published NSI constraints. For $m_X^2\gg q^2$, the interaction approaches the contact limit and can be described by a constant non-standard interaction (NSI) coefficient. For $m_X\lesssim q$, the propagator modifies the recoil spectrum. Constraints obtained in these regimes cannot be directly compared without accounting for their momentum dependence \cite{FarzanTortola2018,Coloma2020,Coloma2021}.

Scalar mediators produce different scattering signatures. Consider a scalar field $\phi$ coupled to Dirac neutrinos, charged leptons, and quarks:
\begin{equation}
\label{eq:scalar}
\Lagr_\phi=
-\phi\left[
y_\nu\bar\nu_R\nu_L+\mathrm{h.c.}
+\sum_{q=u,d,s,\ldots}y_q\bar q q
+y_e\bar e e
\right].
\end{equation}
This is a low-energy effective interaction with a kinematically accessible right-handed component of the Dirac neutrino already present in the model. The massless approximation requires $m_\nu\ll E_\nu,E_\nu-T_N$ and $m_\nu^2\ll2MT_N$ throughout the retained recoil range, with the outgoing state away from a mass threshold. A heavy inaccessible state cannot be represented by the massless cross section below. Equation~\eqref{eq:scalar} is not by itself an electroweak-gauge-invariant UV completion; a completion may correlate the couplings and introduce additional laboratory, cosmological or consistency constraints. No such completion is assumed here.

Scalar couplings to nucleons depend on hadronic matrix elements:
\begin{equation}
\label{eq:scalar_match}
g_{\phi N}
=
\sum_{q=u,d,s}
y_q\frac{m_Nf_{Tq}^{(N)}}{m_q}
+
\frac{2}{27}f_{TG}^{(N)}m_N
\sum_{Q=c,b,t}\frac{y_Q}{m_Q},
\qquad
Q_\phi=Zg_{\phi p}+(A-Z)g_{\phi n},
\end{equation}
where
$f_{TG}^{(N)}=1-\sum_{q=u,d,s}f_{Tq}^{(N)}$.
Unlike vector interactions, scalar nuclear charges cannot be obtained through valence-quark counting \cite{Hoferichter2020,Demirci2024,Papoulias2018,Dent2017Scalar}.

For an ultrarelativistic incoming neutrino and an effectively massless outgoing right-handed state, the leading scalar contribution is
\begin{equation}
\label{eq:scalar_xsec}
\frac{d\sigma_\phi}{dT_N}
\simeq
\frac{|y_\nu|^2Q_\phi^2M^2T_N}
{4\pi E_\nu^2(2MT_N+m_\phi^2)^2}
F_\phi^2(q^2).
\end{equation}
This expression assumes coherent scattering and neglects higher-order recoil corrections. The factor of $T_N$ arises from the chirality-changing neutrino interaction. Unlike the flavor-diagonal vector contribution, this scalar channel does not interfere with the leading SM neutral-current amplitude. The result depends on the assumed neutrino chirality and the accessible final states. For a single initially left-handed massless neutrino, the scalar leptonic spin sum is $|\bar u_R(k')u_L(k)|^2=2k\cdot k'=2MT_N$. The spin-averaged coherent nuclear scalar factor is $4M^2$ at leading recoil order. Thus
\begin{equation*}
\overline{|\mathcal M_\phi|^2}\simeq
\frac{8|y_\nu|^2Q_\phi^2 M^3T_N}{(2MT_N+m_\phi^2)^2}F_\phi^2(q^2),\qquad
\frac{d\sigma}{dT_N}=\frac{\overline{|\mathcal M|^2}}{32\pi ME_\nu^2},
\tag{9a}\label{eq:scalar_trace}
\end{equation*}
which reproduces Eq.~\eqref{eq:scalar_xsec} without averaging over an unpopulated initial right-handed helicity. The SM and scalar amplitudes end in orthogonal left- and right-handed massless states, respectively, so their inclusive interference vanishes. Nonzero masses, mixing with additional accessible states, or different scalar operators would require a new interference calculation. The statement is therefore specific to S1's chiral assumptions.

\begin{figure}[!htbp]
\centering
\includegraphics[width=.83\linewidth]{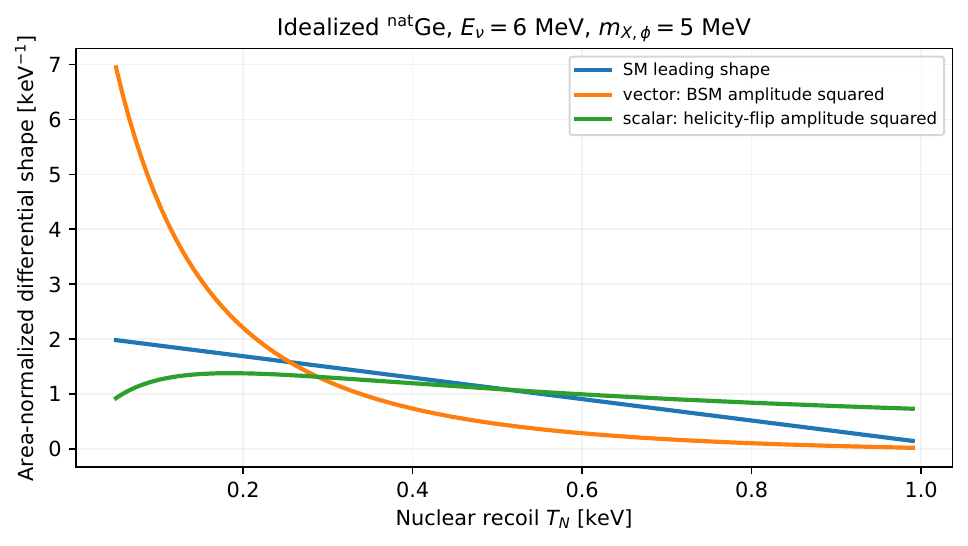}
\caption{Area-normalized recoil spectra for monoenergetic $6\,\mathrm{MeV}$ neutrinos scattering on an idealized germanium nucleus. Both mediators have a mass of $5\,\mathrm{MeV}$. The vector curve represents the squared BSM amplitude without SM interference, while the scalar curve follows Eq.~\eqref{eq:scalar_xsec}. The recoil interval is $0.05$--$0.99\,\mathrm{keV}$. Detector effects and flux integration are not included.}
\label{fig:shapes}
\end{figure}

Scalar-field dark matter can also be tested through its coupling to photons in precision interferometry \cite{Serao2026ScalarInterferometry}. This probes a different interaction from the neutrino--nucleus scalar interaction in Eq.~\eqref{eq:scalar}. The distinct recoil dependence provides a way to discriminate scalar and vector interactions. Vector exchange can interfere with the SM and produces a propagator-dependent enhancement at low momentum transfer. Scalar exchange introduces an additional recoil-energy factor. Figure~\ref{fig:shapes} compares the resulting spectral shapes for an idealized germanium target. The curves illustrate their kinematic differences without detector or flux effects.

Axial-vector and pseudoscalar interactions probe nuclear spin and momentum-dependent responses. They generally lack the full coherent enhancement of vector interactions and require dedicated nuclear calculations \cite{Hoferichter2020,Li2026EFT}. Neutrino magnetic moments provide another source of low-energy recoil enhancement. Their dipole structure differs from scalar and vector exchange. Borexino constrained the effective neutrino magnetic moment using solar-neutrino data \cite{Borexino2017Magnetic}, while XENON1T highlighted the impact of electron-recoil backgrounds \cite{XENON1T2020}. At low deposited energies, atomic binding and ionization effects must also be included \cite{Chen2014Atomic}.

Table~\ref{tab:dictionary} summarizes the main interaction structures and the assumptions required to compare their experimental signatures.

\begin{table}[ht]
\centering\small
\caption{Phenomenological interactions and their leading experimental signatures.}
\label{tab:dictionary}
\begin{tabularx}{\textwidth}{@{}p{0.18\textwidth}X X@{}}
\toprule
Structure & Leading feature & Relevant assumptions \\
\midrule
Flavor-diagonal vector &
SM interference and momentum-dependent weak charge &
Relative sign, flavor charges, quark couplings, mediator mass \\
Flavor-changing vector &
Incoherent contributions from different final flavors &
Source flavor composition and accessible final states \\
Scalar &
Distinct recoil dependence and helicity structure &
Neutrino chirality and hadronic matrix elements \\
Axial/\allowbreak pseudoscalar &
Spin- and momentum-dependent nuclear responses &
Nuclear spin structure and detector acceptance \\
Neutrino magnetic moment &
Enhanced low-energy electron recoils &
Flavor composition and atomic response \\
\bottomrule
\end{tabularx}
\end{table}

For the comparison, we consider three representative interaction models. The first, V1, assumes a vector mediator with universal diagonal neutrino couplings and equal up- and down-quark couplings. Its nuclear charge reduces to $Q_X=3Ag_q$ at zero momentum transfer. The second, V2, allows independent quark couplings and flavor-dependent neutrino interactions. It permits cancellations and degeneracies between nuclear targets. The third, S1, describes chirality-changing scalar exchange with the interaction defined in Eq.~\eqref{eq:scalar}. Unlike V1, S1 as written contains no coupling to a dark-matter field. It predicts a neutrino-scattering contribution, but no dark-matter scattering rate; such a rate would require an additional specified interaction.

These cases distinguish the coupling combinations probed by nuclear recoils, electron recoils, and dark-matter scattering. They do not fix a unique UV completion. Specific gauge models impose further constraints. For example, $U(1)_{B-L}$ requires a consistent fermion content, while $U(1)_{L_\mu-L_\tau}$ can acquire electron and quark interactions through kinetic mixing \cite{Coloma2021,DeRomeri2024Vectors}. Such relations must be specified before applying additional experimental bounds.

Table~\ref{tab:cases} lists their coupling assumptions and the observables they affect.

\begin{table}[t]
\centering\small
\caption{Representative interaction models and the coupling combinations probed by elastic scattering.}
\label{tab:cases}
\begin{tabularx}{\textwidth}{@{}p{.09\textwidth}p{.30\textwidth}X@{}}
\toprule
Case & Interaction hypothesis & Experimental sensitivity \\
\midrule
V1 &
Universal $g_\nu$, $g_u=g_d=g_q$, independent $g_e$ and optional $g_\chi$. &
Nuclear recoils constrain $g_\nu g_q$; electron recoils constrain $g_\nu g_e$; dark-matter recoils constrain $g_\chi g_q$. Individual couplings remain degenerate. \\
V2 &
Independent $g_u$, $g_d$, and diagonal $g_\nu^{\alpha\alpha}$; off-diagonal couplings vanish. &
Different targets and neutrino sources constrain distinct coupling combinations. Nuclear-charge cancellations and nearly parallel blind directions can persist. \\
S1 &
Scalar $\phi$ with Eq.~\eqref{eq:scalar}, accessible massless $\nu_R$, and specified scalar nucleon charges. &
Nuclear recoils constrain $|y_\nu Q_\phi|^2$ with distinct recoil dependence and no leading SM vector interference under the stated chiral assumptions. No dark-matter coupling is defined. \\
\bottomrule
\end{tabularx}
\end{table}

\section{Experimental Results}
\label{sec:exp}

COHERENT first observed coherent elastic neutrino--nucleus scattering (\cevns) using a CsI detector and a stopped-pion neutrino source \cite{Akimov2017}. Subsequent measurements with argon and CsI tested the Standard Model (SM) across different nuclear targets \cite{Akimov2021Argon,Akimov2022CsI}. The source timing separates prompt muon neutrinos from delayed electron neutrinos and muon antineutrinos, providing sensitivity to flavor-dependent interactions.

Germanium measurements have further improved the precision of this program. In 2025, COHERENT Ge-Mini reported $3.9\sigma$ evidence for \cevns\ with a threshold of $1.5\,\mathrm{keV_{ee}}$ \cite{COHERENT2025Ge}. The 2026 analysis lowered the threshold to $0.5\,\mathrm{keV_{ee}}$, corresponding to approximately $2.5\,\mathrm{keV_{nr}}$ under the adopted ionization-yield model. It reported $124^{+14}_{-12}$ events, consistent with the 126 expected from the SM. The measured flux-averaged cross section was $0.98\pm0.10\,(\mathrm{stat})\pm0.10\,(\mathrm{syst})$ times the SM prediction \cite{COHERENT2026Ge}. These results constrain light-mediator interactions through their recoil-energy and timing distributions. Their interpretation depends on flux normalization, detector calibration, and nuclear response \cite{Cadeddu2021,Coloma2020,Carey2026Ar,DeRomeri2023Combined,AtzoriCorona2022Light}.

Reactor experiments probe lower neutrino energies and smaller nuclear recoils. Earlier CONUS and Dresden-II studies established the importance of germanium quenching uncertainties in light-mediator searches \cite{Bonet2022CONUS,AtzoriCorona2022Dresden}. CONUS+ observed reactor antineutrino \cevns\ in 2025 \cite{Ackermann2025}. Its three detectors operated with ionization thresholds of $160$, $170$, and $180\,\mathrm{eV_{ee}}$. These thresholds refer to electron-equivalent energies, not nuclear recoil energies. Their conversion requires a quenching model \cite{LindhardScharff1961}. CONUS+ primarily probes electron-antineutrino interactions and provides constraints complementary to stopped-pion measurements. The resulting bounds depend on reactor fluxes, quenching, detector response, and the assumed mediator structure \cite{DeRomeri2025,Chattaraj2025,AtzoriCorona2025Reactor}.

Dark-matter experiments provide an independent probe of solar-neutrino interactions. PandaX-4T and XENONnT reported the first indications of solar $^8$B neutrinos through nuclear recoils in 2024 \cite{Bo2024PandaX,Aprile2024XENON}. PandaX-4T reached a significance of $2.64\sigma$ and a nuclear-recoil threshold of approximately $0.33\,\mathrm{keV_{nr}}$ using complementary charge-and-light and ionization-only samples \cite{Bo2024PandaX}. In 2026, LZ and XENONnT reported significances of $4.5\sigma$ and $3.3\sigma$, respectively \cite{Akerib2026LZ,Aprile2026XENON}. The LZ low-mass analysis used an exposure of $5.7\,\mathrm{tonne\cdot yr}$ and a nuclear-recoil interval of $1$--$6\,\mathrm{keV}$. These measurements probe light mediators in the same recoil region relevant to dark-matter searches. A study of fermionic and scalar dark matter with matched xenon responses found different solar-capture histories and neutrino source spectra \cite{Serao2026SolarCapture}. However, neutrino fluxes, nuclear responses, and detector backgrounds remain correlated with the inferred interaction strength.

Electron recoils provide access to different mediator couplings. In 2026, XENONnT reported a $5.0\sigma$ observation of solar-neutrino--electron scattering using $2.46\,\mathrm{tonne\cdot yr}$ of exposure and electron-recoil energies between $1$ and $140\,\mathrm{keV}$ \cite{Aprile2026Electron}. Borexino provides complementary measurements of solar-neutrino fluxes and flavor conversion \cite{Borexino2018,Borexino2022,OrebiGann2021}. These results constrain interactions involving electrons, while nuclear recoils probe combinations of quark couplings. Their interpretation requires the solar-neutrino flavor composition and the electron response of the detector \cite{Chen2014Atomic}. Background uncertainties remain important, as illustrated by the earlier XENON1T electron-recoil excess \cite{XENON1T2020}.

For free electrons, the leading SM neutrino--electron scattering cross section is
\begin{equation}
\label{eq:nue}
\frac{d\sigma_{\nu e}}{dT_e}
=
\frac{2G_F^2m_e}{\pi}
\left[
g_L^2+
g_R^2\left(1-\frac{T_e}{E_\nu}\right)^2
-g_Lg_R\frac{m_eT_e}{E_\nu^2}
\right],
\end{equation}
where $g_R=\sin^2\theta_W$ and $g_L=\sin^2\theta_W+\tfrac12$ for $\nu_e e$ scattering, including the charged-current contribution. For $\nu_{\mu,\tau}e$, $g_L=\sin^2\theta_W-\tfrac12$. The left- and right-handed terms are interchanged for antineutrinos.

A flavor-diagonal vector mediator modifies these couplings through
\begin{equation}
\label{eq:electronshift}
\delta g_L=\delta g_R=
\frac{g_\nu^{\alpha\alpha}g_e}
{2\sqrt{2}G_F(2m_eT_e+m_X^2)}.
\end{equation}
The sign depends on the coupling product. Electron scattering probes $q_e^2\simeq2m_eT_e$, while nuclear scattering probes $q_N^2\simeq2MT_N$. The two channels therefore sample different mediator propagators. Solar-neutrino flavor conversion and atomic binding further modify the electron-recoil spectrum. Constraints on $g_\nu g_e$ and $g_\nu g_q$ can be combined only after specifying the relations between electron and quark couplings.

Table~\ref{tab:numbers} summarizes the main experimental results and their relevance to light-mediator searches.

\begin{table}[t]
\centering
\small
\caption{Selected experimental results relevant to light-mediator searches. Thresholds, exposures, and significances refer to different analyses and cannot be directly compared as measures of mediator sensitivity.}
\label{tab:numbers}
\begin{tabularx}{\textwidth}{@{}p{.23\textwidth}p{.35\textwidth}X@{}}
\toprule
Experiment & Main result & Relevant information \\
\midrule
COHERENT CsI/Ar
\cite{Akimov2017,Akimov2021Argon,Akimov2022CsI}
&
First \cevns\ observation and subsequent target measurements
&
Flavor timing and nuclear-target dependence
\\

COHERENT Ge, 2025
\cite{COHERENT2025Ge}
&
$3.9\sigma$ evidence; $1.5\,\mathrm{keV_{ee}}$ threshold
&
First Ge-Mini evidence
\\

COHERENT Ge, 2026
\cite{COHERENT2026Ge}
&
$124^{+14}_{-12}$ events;
$\sigma/\sigma_{\rm SM}
=0.98\pm0.10_{\rm stat}\pm0.10_{\rm syst}$
&
$0.5\,\mathrm{keV_{ee}}$ threshold; energy and timing information
\\

CONUS+, 2025
\cite{Ackermann2025}
&
Reactor \cevns\ observation; minimum threshold $160\,\mathrm{eV_{ee}}$
&
Electron-antineutrino interactions and Ge quenching
\\

PandaX-4T, 2024
\cite{Bo2024PandaX}
&
$2.64\sigma$ solar-$^8$B indication
&
Low-threshold xenon nuclear recoils
\\

LZ, 2026
\cite{Akerib2026LZ}
&
$4.5\sigma$ solar-$^8$B evidence;
$5.7\,\mathrm{t\,yr}$
&
Neutrino nuclear recoils and dark-matter search overlap
\\

XENONnT, 2026
\cite{Aprile2026XENON}
&
$3.3\sigma$ solar-$^8$B nuclear-recoil result
&
Independent xenon measurement
\\

XENONnT, 2026
\cite{Aprile2026Electron}
&
$5.0\sigma$ neutrino--electron signal;
$2.46\,\mathrm{t\,yr}$
&
Electron couplings and solar-neutrino flavor composition
\\
\bottomrule
\end{tabularx}
\end{table}

These experiments probe different neutrino sources, momentum transfers, nuclear targets, and fermion couplings. Their combination can resolve mediator-parameter degeneracies, but experimental precision alone does not guarantee complementarity. Reliable constraints require consistent flux models, recoil responses, coupling conventions, and statistical treatments. The following section examines how these ingredients determine the parameters actually constrained by each measurement.

\section{What Is Actually Constrained?}
\label{sec:constraints}

Experimental constraints on light mediators depend on the interaction model, neutrino flux, target response, and statistical treatment. The expected event rate in reconstructed-energy bin $i$ of experiment $a$ is
\begin{equation}
\label{eq:rate}
\mu_{ai}(\bm\theta,\bm\eta)
=
b_{ai}(\bm\eta)
+
\mathcal{E}_a
\sum_{s,\alpha}
\int dE_\nu\,dT\,
\Phi_{s\alpha}(E_\nu;\bm\eta)
\frac{d\sigma_{a\alpha}}{dT}
(E_\nu,T;\bm\theta,\bm\eta)
\mathcal{R}_{ai}(T;\bm\eta),
\end{equation}
where $\bm\theta$ denotes the mediator parameters and $\bm\eta$ the nuisance parameters. The background contribution is $b_{ai}$, $\mathcal{E}_a$ is the exposure, and $\mathcal{R}_{ai}$ includes detector response and selection efficiency. The index $s$ labels neutrino sources. Equation~\eqref{eq:rate} shows that experimental limits constrain the full event prediction, not individual mediator couplings.

\begin{figure}[!htbp]
\centering
\includegraphics[width=\linewidth]{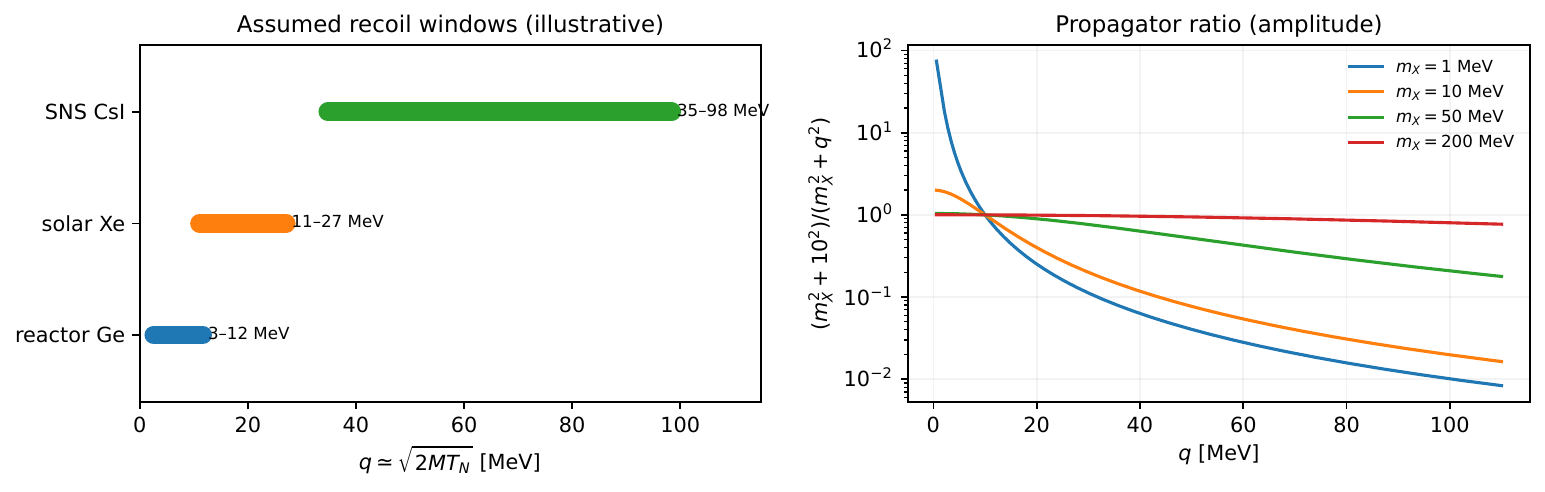}
\par\vspace{5pt}
\includegraphics[width=\linewidth]{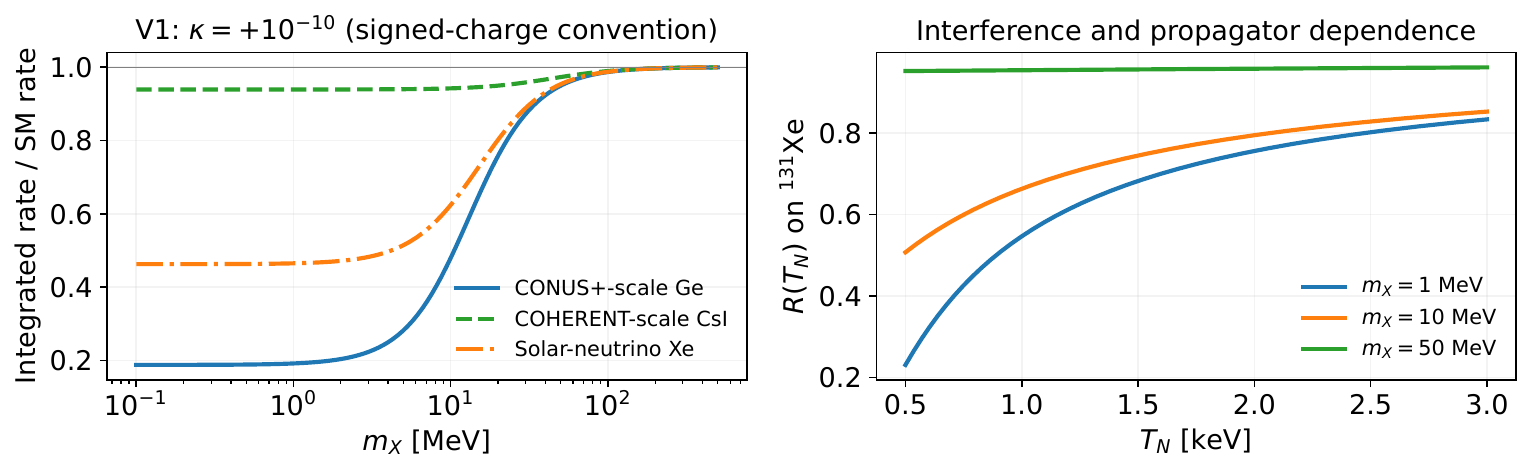}
\caption{Upper panels: the original illustrative momentum windows (left) and propagators relative to $q=10\,\mathrm{MeV}$ (right). Lower panels: flux-integrated V1 rate ratios versus mediator mass and differential Xe ratios for $\kappa=+10^{-10}$. The lower panels use the separately specified windows and inputs of Table~\ref{tab:v1_rates}, including distinct proton and neutron form factors for CsI. All windows are illustrative rather than experimental acceptance ranges.}
\label{fig:momentum}
\end{figure}

Momentum transfer determines the mediator propagator. For nonrelativistic nuclear recoils, $q\simeq\sqrt{2MT_N}$. Different targets and recoil thresholds therefore probe different momentum ranges. Figure~\ref{fig:momentum} compares representative recoil intervals: $0.05$--$1.0\,\mathrm{keV}$ for reactor-scale Ge, $0.5$--$3\,\mathrm{keV}$ for solar-neutrino Xe, and $5$--$40\,\mathrm{keV}$ for stopped-pion CsI. These intervals correspond approximately to $q=3$--$12$, $11$--$27$, and $35$--$99\,\mathrm{MeV}$, respectively. They illustrate the relevant kinematic regimes rather than the measured acceptance of individual experiments.

For $m_X^2\gg q^2$, the mediator propagator approaches the contact limit:
\begin{equation}
\frac{1}{q^2+m_X^2}
=
\frac{1}{m_X^2}
\left[
1-\frac{q^2}{m_X^2}
+\mathcal{O}\left(\frac{q^4}{m_X^4}\right)
\right].
\end{equation}
The interaction can then be described by an approximately constant effective coupling. For $m_X^2\ll q^2$, the amplitude scales as $1/q^2$. Between these regimes, the propagator modifies the recoil spectrum. Experiments probing different momentum ranges can therefore constrain the same coupling product with different sensitivities \cite{Cadeddu2021,Dent2017}. Quantitative comparisons require flux integration and detector-response modeling.

\paragraph*{A common V1 response calculation.}
To compare the three source regimes at fixed interaction parameters, we set
$g_u=g_d=g_q$ and $g_\nu^{\alpha\alpha}=g_\nu$ for all active flavors, and define
$\kappa=g_\nu g_q$. With the signed weak charge specified in Section~\ref{sec:model},
Eq.~\eqref{eq:shift} gives
\begin{equation*}
\begin{gathered}
\Delta Q_W=\frac{3\sqrt{2}A\kappa}{G_F(q^2+m_X^2)},\qquad
d(T_N,m_X)=\frac{3\sqrt{2}A}{G_FQ_W(2MT_N+m_X^2)},\\
R(T_N)=\left|1+\frac{\Delta Q_W}{Q_W}\right|^2
       =[1+\kappa d(T_N,m_X)]^2.
\end{gathered}
\tag{13a}\label{eq:v1_ratio}
\end{equation*}
Here $\kappa$ is real. Since $Q_W<0$ for the nuclei considered, positive $\kappa$
initially suppresses the SM rate. The quadratic term is retained throughout the
finite-coupling comparison; sufficiently large positive couplings can pass through
an amplitude zero and enhance the rate. Changing the sign convention for $Q_W$
requires changing the sign of the shift at the same time.

Equation~\eqref{eq:v1_ratio} assumes a common proton and neutron form factor.
At the larger CsI momentum transfers, we instead use
\begin{equation*}
\begin{split}
\mathcal Q_W(q)&=(1-4\sin^2\theta_W)ZF_p(q)-(A-Z)F_n(q),\\
d(T_N,m_X)&=\frac{3\sqrt{2}\,[ZF_p(q)+(A-Z)F_n(q)]}
 {G_F(q^2+m_X^2)\mathcal Q_W(q)},\qquad
R(T_N)=[1+\kappa d(T_N,m_X)]^2.
\end{split}
\tag{13b}\label{eq:v1_ratio_ff}
\end{equation*}
The SM spectrum in this expression contains $\mathcal Q_W^2(q)$ rather than
$Q_W^2F_W^2(q)$. Thus the common-form-factor cancellation is not imposed on CsI.
We use Helm form factors \cite{Helm1956},
$F(q)=3j_1(qR_1)/(qR_1)\exp[-(qs)^2/2]$, with $s=0.9\,\mathrm{fm}$ and
$\hbar c=0.1973269804\,\mathrm{GeV\,fm}$ when evaluating their arguments.
For Cs and I separately, $R_1^2=5(R_{\rm rms}^2-3s^2)/3$ with
$(R_p,R_n)_{\rm Cs}=(4.821,5.09)\,\mathrm{fm}$ and
$(R_p,R_n)_{\rm I}=(4.766,5.03)\,\mathrm{fm}$, following the inputs of
Ref.~\cite{AtzoriCorona2022Light}. These are fixed nuclear-model inputs, not a
fit of neutron radii. Ge and Xe use a common Helm form factor with
$R_1^2=c^2+7\pi^2a_H^2/3-5s^2$, $c=(1.23A^{1/3}-0.60)\,\mathrm{fm}$ and
$a_H=0.52\,\mathrm{fm}$. Their neutron-skin differences, axial responses and
radiative corrections are neglected in this illustrative calculation.

For each nucleus $a$, let $w_a(T)=\int dE_\nu\,\Phi(E_\nu)
(d\sigma_{{\rm SM},a}/dT)$ denote its flux-averaged spectrum. The integrated
response is
\begin{equation*}
\overline R(m_X,\kappa)=
\frac{\displaystyle\sum_a N_a\int_{T_{\min}}^{T_{\max}}dT\,
 w_a(T)[1+\kappa d_a(T,m_X)]^2}
{\displaystyle\sum_a N_a\int_{T_{\min}}^{T_{\max}}dT\,w_a(T)},
\qquad \frac{\Delta\mathcal N}{\mathcal N_{\rm SM}}=\overline R-1.
\tag{13c}\label{eq:v1_integrated}
\end{equation*}
The relative numbers of nuclei are $N_{\rm Cs}:N_{\rm I}=1:1$; cross sections
are summed before taking the ratio. For Ge and Xe we use single representative
isotopes, $^{73}$Ge and $^{131}$Xe, explicitly replacing the natural isotope
mixtures. All windows are in true nuclear-recoil energy, with unit efficiency,
perfect energy reconstruction and no non-neutrino backgrounds. They represent
source regimes, not the measured acceptance of a particular detector.

The reactor example uses a pure-$^{235}$U spectral surrogate on
$2$--$8\,\mathrm{MeV}$ \cite{Huber2011},
\[
\begin{gathered}
\Phi(E_\nu)\propto\exp\!\left[\sum_{k=0}^{5}a_k(E_\nu/\mathrm{MeV})^k\right],\\
\bm a=(4.367,-4.577,2.100,-0.5294,0.06186,-0.002777).
\end{gathered}
\]
This parametrization and truncation define the benchmark;
they do not reproduce the CONUS+ reactor fuel mixture or flux uncertainties.
The stopped-pion spectrum contains equal integrated numbers of prompt
$\nu_\mu$ at $E_\pi=(m_\pi^2-m_\mu^2)/(2m_\pi)=29.79\,\mathrm{MeV}$ and delayed
$\nu_e,\bar\nu_\mu$. Their normalized continuous distributions are
$f_{\nu_e}=12x^2(1-x)/E_{\max}$ and
$f_{\bar\nu_\mu}=2x^2(3-2x)/E_{\max}$, where
$x=E_\nu/E_{\max}$ and $E_{\max}=m_\mu/2=52.83\,\mathrm{MeV}$
\cite{AtzoriCorona2022Light}; all emission times are accepted.
The solar calculation uses the central $^8$B spectrum tabulated by
Winter et al.~\cite{Winter2006}, linearly interpolated and normalized to unit
integral. The total active flux suffices because the assumed coupling is flavor
universal. Flux normalizations and exposures cancel in Eq.~\eqref{eq:v1_integrated}.

We fix $\kappa=+10^{-10}$, $\sin^2\theta_W=0.23857$,
$G_F=1.1663787\times10^{-5}\,\mathrm{GeV}^{-2}$ and $M=A\times0.9315\,\mathrm{GeV}$.
All propagators are evaluated in $\mathrm{GeV}^2$, using
$T[\mathrm{GeV}]=10^{-6}T[\mathrm{keV}]$ and
$m_X[\mathrm{GeV}]=10^{-3}m_X[\mathrm{MeV}]$.
The neutrino integrals enforce the exact massless elastic threshold
$E_{\min}=(T+\sqrt{T^2+2MT})/2$, while retaining the leading recoil kernel of
Eq.~\eqref{eq:sm}. Adding its subleading $-T/E_\nu$ term changes the integrated
SM rates by less than $0.1\%$ in these examples; this small check does not quantify
the omitted detector, flux or nuclear uncertainties.

\begin{table}[t]
\centering\small
\caption{Predicted V1 rate changes for $\kappa=+10^{-10}$. Columns headed by
$m_X$ give $100(\overline R-1)$ after flux and recoil integration. The Ge window
illustrates the CONUS+ source regime, CsI the COHERENT regime, and the common Xe
window the solar-neutrino regime of LZ, XENONnT and PandaX-4T. The latter three
are not assigned different predictions without their distinct detector responses.
These are model responses, not sensitivity estimates or exclusions.}
\label{tab:v1_rates}
\begin{tabular}{@{}lccrrr@{}}
\toprule
Source / target & $T$ [keV] & $q$ [MeV] & $1$ MeV & $10$ MeV & $50$ MeV \\
\midrule
Reactor / Ge & 0.7--1.5 & 9.8--14.3 & $-80.77\%$ & $-52.03\%$ & $-5.06\%$ \\
Stopped-pion / CsI & 5--40 & 34.4--99.6 & $-6.07\%$ & $-5.77\%$ & $-2.70\%$ \\
Solar $^8$B / Xe & 0.5--3 & 11.0--27.1 & $-53.47\%$ & $-37.59\%$ & $-4.65\%$ \\
\bottomrule
\end{tabular}

\end{table}

Table~\ref{tab:v1_rates} and the lower panels of Figure~\ref{fig:momentum}
show why a common coupling product produces different fractional responses.
At $m_X=1\,\mathrm{MeV}$ the differential ratios across the adopted windows are
$0.091$--$0.452$ for Ge, $0.896$--$0.986$ for the summed CsI spectrum, and
$0.231$--$0.833$ for Xe. A rate ratio is consequently not represented by a
single arbitrarily chosen momentum. At larger mediator mass the response
approaches the contact scaling, although $50\,\mathrm{MeV}$ is not asymptotically
heavy relative to the entire CsI momentum window.

\paragraph*{Relation to published exclusions.}
In a universal vector model written with the neutrino current
$\bar\nu\gamma^\mu P_L\nu$ and equal neutrino and quark couplings $g'$, the
nuclear-scattering sector maps to V1 through $\kappa=(g')^2\geq0$.
It does not cover the negative-product branch. The COHERENT universal-vector
study in Ref.~\cite{AtzoriCorona2022Light} reports $2\sigma$ ($95.45\%$)
regions, including disconnected solutions from destructive interference. Its published low-mass boundaries on $g'$ for Ar, CsI and their combination, respectively, are $3.91\times10^{-5}$, $2.36\times10^{-5}$ and $2.07\times10^{-5}$. Squaring these gives the corresponding V1 products $\kappa=1.53\times10^{-9}$, $5.57\times10^{-10}$ and $4.28\times10^{-10}$. These are conversions of the historical $95.45\%$ low-mass boundaries in that study, not a new fit, exact finite-mass interpolations, or bounds on the negative-$\kappa$ branch.
The XENONnT solar-neutrino analysis reports a $90\%$ upper-limit curve for
the universal vector case with its dark-matter contribution fixed to zero
\cite{Aprile2026XENON}. Its flavor-universal coupling makes active-flavor
redistribution irrelevant to the nuclear rate. CONUS+ interpretations also
include neutrino--electron events in some models \cite{Chattaraj2025}; those
constraints require an electron-coupling prescription beyond the independent
$g_e$ of V1. Public COHERENT data and XENONnT recoil-response templates are
available \cite{COHERENTRelease,XENONRelease2026}, but their availability does
not turn different confidence prescriptions or nuisance treatments into a
common likelihood. No common exclusion envelope is extracted or combined
here: doing so would require reproducing each published SM fit, retaining its
interference branches and applying a common statistical prescription. The
quantitative comparison in Table~\ref{tab:v1_rates} is restricted to predictions.

Neutrino flavor provides an additional source of information. Reactor experiments primarily probe electron antineutrinos. Stopped-pion sources provide prompt muon neutrinos and delayed electron neutrinos and muon antineutrinos. Solar neutrinos reach detectors as energy-dependent flavor mixtures. Flavor-dependent couplings $g_\nu^{ee}$, $g_\nu^{\mu\mu}$, and $g_\nu^{\tau\tau}$ therefore affect these measurements differently. CEvNS can also constrain degeneracies present in neutrino-oscillation analyses \cite{FarzanTortola2018,Coloma2017,ColomaDenton2017,Coloma2020}. However, oscillations probe forward scattering, while recoil measurements involve finite momentum transfer. Combining them requires a consistent mediator model and charge assignment \cite{Coloma2021,Wolfenstein1978,Li2025Combined}. Other modifications of neutrino propagation arise in curved-spacetime field mixing and torsion backgrounds \cite{Capolupo:2020wlx, Capolupo:2021cnt, Capolupo:2023fao, Serao2023CurvedMixing,Serao2024Torsion,Serao2026SpinTorsion}. These mechanisms must be distinguished from light-mediator matter potentials.

Nuclear composition introduces further degeneracies. The vector charge in Eq.~\eqref{eq:qx} depends on both $g_u$ and $g_d$. Different coupling combinations can suppress scattering in one nucleus while remaining accessible in another. Ge, Ar, CsI, and Xe therefore probe different directions in coupling space \cite{Akimov2021Argon,COHERENT2026Ge,Ackermann2025}. Nuclear form factors and neutron distributions introduce additional uncertainties, particularly at higher momentum transfer \cite{TestingCoherence2024,Canas2020,AbdelKhaleq2024}. Atomic parity violation provides complementary information on electroweak parameters and neutron distributions \cite{CadedduDordei2019}, although constraints on new mediators require the appropriate parity-violating interactions.

The role of nuclear composition can be quantified using the vector charge per nucleon:
\begin{equation}
\label{eq:charge_linear}
\frac{Q_X}{A}
=
\left(1+\frac{Z}{A}\right)g_u
+
\left(2-\frac{Z}{A}\right)g_d.
\end{equation}
For two targets, the coefficients of $g_u$ and $g_d$ define a $2\times2$ response matrix. Its determinant is proportional to the difference between their proton fractions $Z/A$. Identical proton fractions therefore produce an exact degeneracy.

For representative isotopes $^{73}\mathrm{Ge}$ and $^{131}\mathrm{Xe}$, the matrix is
\begin{equation}
\label{eq:matrix}
C=
\begin{pmatrix}
105/73 & 114/73\\
185/131 & 208/131
\end{pmatrix}.
\end{equation}
Its 2-norm condition number is $\kappa_2(C)\simeq115.04$. The large value indicates that small uncertainties in the inferred nuclear charges can produce much larger uncertainties in the reconstructed quark couplings. The two targets therefore provide nearly parallel directions in coupling space.

\begin{figure}[!htbp]
\centering
\includegraphics[width=.86\linewidth]{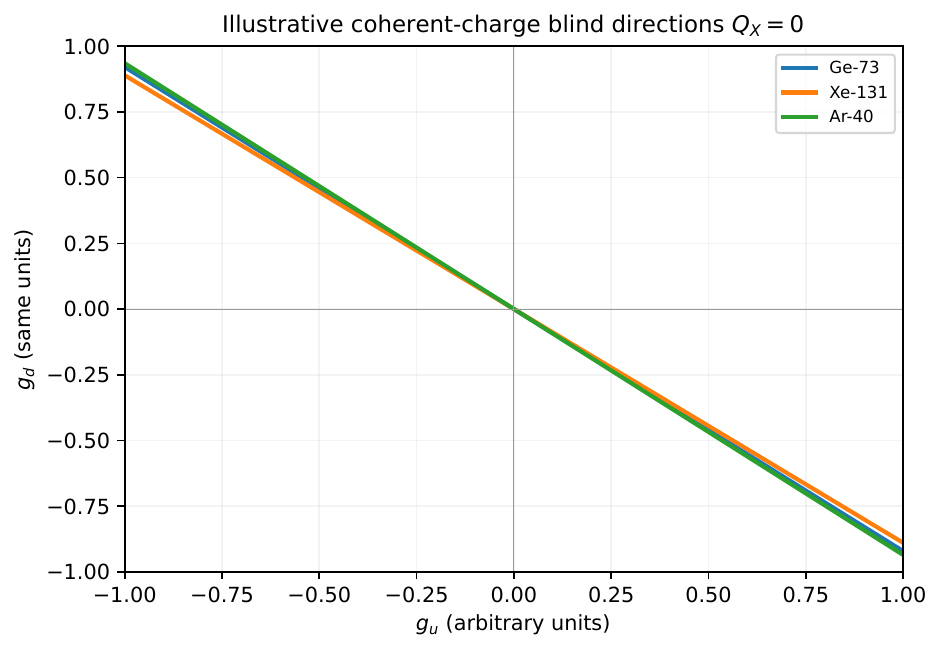}
\par\vspace{4pt}
\includegraphics[width=.86\linewidth]{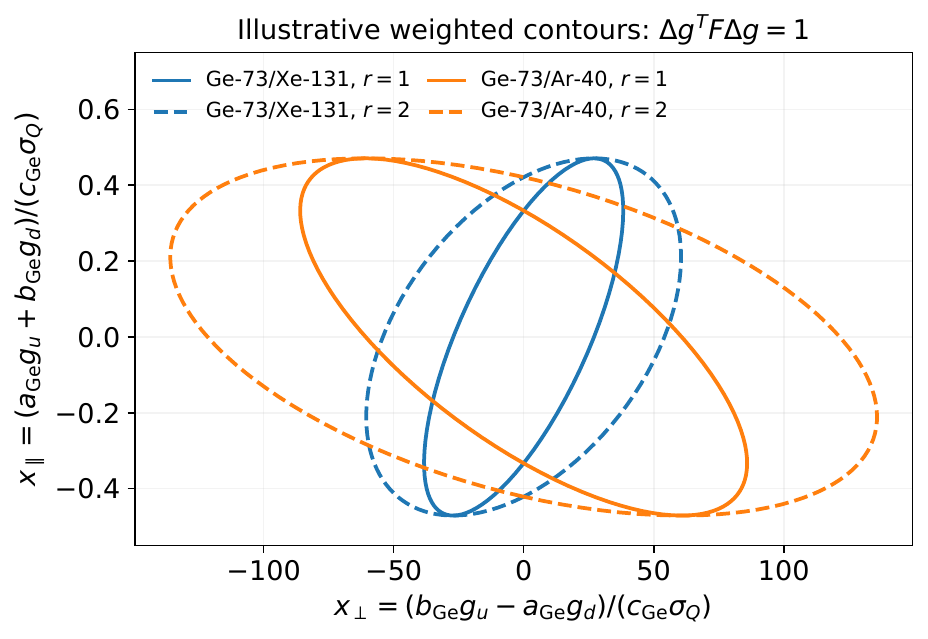}
\caption{Upper panel: the original zero-charge conditions $Q_X=0$ for $^{73}$Ge, $^{131}$Xe and $^{40}$Ar in the $(g_u,g_d)$ plane, at fixed neutrino couplings and common normalized proton/neutron form factors. Lower panel: illustrative contours $\Delta\bm g^TF\Delta\bm g=1$ for the two target pairs and errors in Table~\ref{tab:weighted_charge}, centered at zero. The common coordinates resolve directions perpendicular and parallel to $\bm c_{\rm Ge}=(a_{\rm Ge},b_{\rm Ge})$, where $c_{\rm Ge}=|\bm c_{\rm Ge}|$; the axes have different scales. The contours do not represent confidence regions obtained from data.}
\label{fig:blind}
\end{figure}

Figure~\ref{fig:blind} shows the corresponding zero-charge conditions. Although Ge and Xe have different nuclear compositions, their blind directions remain close. Adding a second target does not necessarily resolve the degeneracy. The result assumes fixed neutrino couplings and zero-momentum nuclear charges. Realistic isotope mixtures, experimental uncertainties, and interference effects modify the inference.

\paragraph*{Weighted information and its limitations.}
The geometric condition number does not encode experimental precision.
To separate the two, suppose that inferred charge-per-nucleon measurements
$\bm y$ have Gaussian covariance $\Sigma$ and expectation $C\bm g$, where
$\bm g=(g_u,g_d)^T$. This assumes a fixed neutrino coupling, mediator mass and
local interference branch. Expanding their likelihood gives
\begin{equation*}
\begin{gathered}
-2\ln L=(\bm y-C\bm g)^T\Sigma^{-1}(\bm y-C\bm g)+\text{constant},\\
F_{ij}=-\left\langle\frac{\partial^2\ln L}{\partial g_i\partial g_j}\right\rangle
=(C^T\Sigma^{-1}C)_{ij}.
\end{gathered}
\tag{15a}\label{eq:charge_fisher}
\end{equation*}
For independent uncertainties $\sigma_1,\sigma_2$, set $a_t=1+Z_t/A_t$ and
$b_t=2-Z_t/A_t$. Then
\begin{equation*}
F=\begin{pmatrix}
\sum_t a_t^2/\sigma_t^2 & \sum_t a_tb_t/\sigma_t^2\\
\sum_t a_tb_t/\sigma_t^2 & \sum_t b_t^2/\sigma_t^2
\end{pmatrix},\qquad
\det F=\frac{9[(Z_1/A_1)-(Z_2/A_2)]^2}{\sigma_1^2\sigma_2^2}.
\tag{15b}\label{eq:charge_fisher_explicit}
\end{equation*}
For correlated errors, a factorization $\Sigma=LL^T$ gives the whitened response
$W=L^{-1}C$ and $F=W^TW$. If $F_{11}=u$, $F_{12}=v$ and $F_{22}=w$, its
eigenvalues and normalized eigenvectors are
\begin{equation*}
\begin{gathered}
\lambda_\pm=\frac{u+w\pm\sqrt{(u-w)^2+4v^2}}{2},\qquad
\bm v_\pm=\frac{(v,\lambda_\pm-u)^T}{\sqrt{v^2+(\lambda_\pm-u)^2}},\\
\kappa_2(W)=\sqrt{\frac{\lambda_+}{\lambda_-}},\quad
\kappa_2(F)=\kappa_2(W)^2.
\end{gathered}
\tag{15c}\label{eq:charge_eigenvalues}
\end{equation*}
The eigenvector formula applies to the nonzero off-diagonal entries used here.

For illustration only, we choose absolute errors $\sigma_1=\sigma_Q=10^{-5}$
and $\sigma_2=r\sigma_Q$, with $r=1$ or $2$ and no correlation. They are errors
on hypothetical measurements of $Q_X/A$, not fractional rate uncertainties or
errors extracted from an experiment. Table~\ref{tab:weighted_charge} gives the
eigenvalues of $\overline F=\sigma_Q^2F$, its weak direction, and the condition
number of $W$. The strong eigenvector is
$\bm v_+=(-v_{-,d},v_{-,u})^T$. For example, when $r=2$ the matrices are
\begin{equation*}
\overline F_{\mathrm{Ge/Xe}}=
\begin{pmatrix}2.56746&2.80677\\2.80677&3.06900\end{pmatrix},\qquad
\overline F_{\mathrm{Ge/Ar}}=
\begin{pmatrix}2.59449&2.80808\\2.80808&3.03936\end{pmatrix}.
\tag{15d}\label{eq:charge_numeric}
\end{equation*}
The scripts retain full precision rather than diagonalizing these rounded entries.

\begin{table}[t]
\centering\small
\caption{Illustrative weighted charge information. Eigenvalues refer to
$\overline F=\sigma_Q^2F$; the first eigenvalue column is in units of $10^{-4}$.
Eigenvectors are given in the $(g_u,g_d)$ basis. These inputs and contours are
not confidence regions inferred from measured recoil spectra.}
\label{tab:weighted_charge}
\begin{tabular}{@{}lccclr@{}}
\toprule
Pair & $r$ & $\lambda_-/10^{-4}$ & $\lambda_+$ & $\bm v_-$ & $\kappa_2(W)$ \\
\midrule
$^{73}$Ge/$^{131}$Xe & 1 & 6.8173 & 9.0223 & $(0.74141,-0.67105)$ & 115.04 \\
$^{73}$Ge/$^{131}$Xe & 2 & 2.7283 & 5.6362 & $(0.73790,-0.67491)$ & 143.73 \\
$^{73}$Ge/$^{40}$Ar & 1 & 1.3539 & 9.0125 & $(0.73291,-0.68032)$ & 258.00 \\
$^{73}$Ge/$^{40}$Ar & 2 & 0.5415 & 5.6338 & $(0.73449,-0.67861)$ & 322.56 \\
\bottomrule
\end{tabular}

\end{table}

The Ge/Ar pair is more ill-conditioned than Ge/Xe at the same precision:
$32/73=0.43836$ is closer to $18/40=0.45000$ than to $54/131=0.41221$.
With equal errors the two condition numbers are $258.00$ and $115.04$;
doubling the uncertainty of the second target increases them to $322.56$ and
$143.73$, respectively. Figure~\ref{fig:blind} also shows the weighted quadratic
contours in a common rotated coupling basis. Their widths encode the assumed
precision as well as the charge geometry. A uniform rescaling of both errors
changes the contour size and Fisher eigenvalues, but leaves the condition number
unchanged; conditioning alone therefore cannot measure absolute sensitivity.

For actual Poisson recoil data, the relevant information is instead
\[
F^{\rm data}_{ij}=\sum_{a,k}\mu_{ak}^{-1}
\frac{\partial\mu_{ak}}{\partial g_i}\frac{\partial\mu_{ak}}{\partial g_j},
\]
with the derivatives of Eq.~\eqref{eq:rate}. Profiling nuisance parameters further gives
$F_{gg}^{\rm prof}=F_{gg}-F_{g\eta}F_{\eta\eta}^{-1}F_{\eta g}$.
The derivative can even vanish at an amplitude cancellation, and disconnected
rate solutions cannot be described by a single Gaussian charge covariance.
Spectral information, flux and detector correlations, isotope mixtures and
interference thus separate the geometric calculation from a realistic likelihood.

Experimental constraints also depend on statistical assumptions. Published bounds may use frequentist confidence intervals, Bayesian credible regions, or expected sensitivity estimates. They may fix individual couplings or profile them alongside experimental uncertainties. These prescriptions are not interchangeable. Correlations between neutrino fluxes, backgrounds, nuclear responses, and detector parameters can also change the inferred bounds. Combined nuclear- and electron-recoil analyses must account for these effects \cite{AtzoriCorona2026}.

Table~\ref{tab:metadata} summarizes the information required for a consistent reinterpretation of light-mediator constraints.

\begin{table}[ht]
\centering
\small
\caption{Information required to compare and reinterpret light-mediator constraints.}
\label{tab:metadata}
\begin{tabularx}{\textwidth}{@{}p{0.25\textwidth}X@{}}
\toprule
Item & Required information \\
\midrule
Interaction model &
Mediator spin, flavor structure, fermion couplings, and charge assignments \\
Parameter definitions &
Coupling products, relative signs, units, and fixed or profiled parameters \\
Mediator regime &
Propagator dependence, mediator mass, and effective-theory validity \\
Neutrino inputs &
Flux spectra, flavor composition, oscillation parameters, and uncertainties \\
Nuclear and atomic response &
Isotope composition, form factors, quenching, and electron binding \\
Detector response &
Thresholds, efficiencies, energy resolution, and background models \\
Statistical treatment &
Likelihood, nuisance parameters, correlations, and confidence conventions \\
Reproducibility &
Public data, response matrices, numerical inputs, and analysis codes \\
\bottomrule
\end{tabularx}
\end{table}

Two limitations follow from this comparison. Different momentum ranges do not necessarily provide independent constraints on mediator couplings. Likewise, different nuclear targets can retain nearly identical blind directions. Reliable combinations require distinct parameter sensitivities and consistent statistical models. Both conditions are essential for interpreting light-mediator searches across neutrino and dark-matter experiments.

\section{Neutrino Backgrounds in Dark-Matter Searches}
\label{sec:fog}

Neutrino-induced nuclear recoils constitute an important background in dark-matter searches, as emphasized by successive xenon-based searches \cite{LZ2023}. Their spectra can resemble those produced by weakly interacting massive particles (WIMPs), reducing discovery sensitivity \cite{Billard2014}. Combining different nuclear targets can improve signal discrimination \cite{Ruppin2014}. However, the neutrino background does not impose an absolute detection limit. The neutrino fog instead describes a regime where sensitivity depends on spectral overlap, exposure, and flux uncertainties \cite{OHare2021}. Light mediators can further modify this regime by changing neutrino scattering rates and recoil spectra \cite{Dent2017}.

Recent measurements by LZ and XENONnT have established solar-neutrino nuclear recoils as observable signals in dark-matter detectors \cite{Akerib2026LZ,Aprile2026XENON}. These measurements also constrain new neutrino interactions. However, such interactions can mimic or obscure dark-matter signals. Their effects must therefore be included when determining dark-matter discovery sensitivity.

For a vector mediator coupling to both neutrinos and dark matter, neutrino scattering depends on $g_\nu g_q$, while dark-matter scattering depends on $g_\chi g_q$. Both processes depend on the mediator mass and detector response. Unknown neutrino couplings can weaken dark-matter constraints, while relations imposed by specific ultraviolet models can reduce these degeneracies \cite{Dent2017,OHare2021}.

We quantify this effect using an analytic likelihood model. Near a reference background, the expected event count in bin $i$ is $\mu_i=b_i+\alpha s_i+\beta n_i$, where $b_i$ is the reference expectation, $s_i$ the dark-matter signal template, and $n_i$ a neutrino-background deformation. The parameters $\alpha$ and $\beta$ control their amplitudes. For independent Poisson bins in the Gaussian-information limit, with an external Gaussian constraint of width $\sigma_\beta$, the Fisher information is
\begin{equation}
\label{eq:fisher}
I_{\alpha\alpha}=\sum_i\frac{s_i^2}{b_i},
\qquad
I_{\alpha\beta}=\sum_i\frac{s_i n_i}{b_i},
\qquad
I_{\beta\beta}=\sum_i\frac{n_i^2}{b_i}+\frac{1}{\sigma_\beta^2},
\qquad
I_{\rm prof}=I_{\alpha\alpha}-\frac{I_{\alpha\beta}^2}{I_{\beta\beta}}.
\end{equation}
Profiling over $\beta$ reduces the information available on the dark-matter amplitude. Defining the weighted spectral overlap
$\rho=I_{\alpha\beta}/\sqrt{I_{\alpha\alpha}I_{\beta\beta}^{\rm data}}$
and the relative external constraint
$\lambda=(\sigma_\beta^2 I_{\beta\beta}^{\rm data})^{-1}$,
the uncertainty increases by
\begin{equation}
\label{eq:degrade}
\frac{\sigma_{\alpha,\rm prof}}{\sigma_{\alpha,\rm fixed}}
=
\left(1-\frac{\rho^2}{1+\lambda}\right)^{-1/2}.
\end{equation}

\begin{figure}[!htbp]
\centering
\includegraphics[width=.82\linewidth]{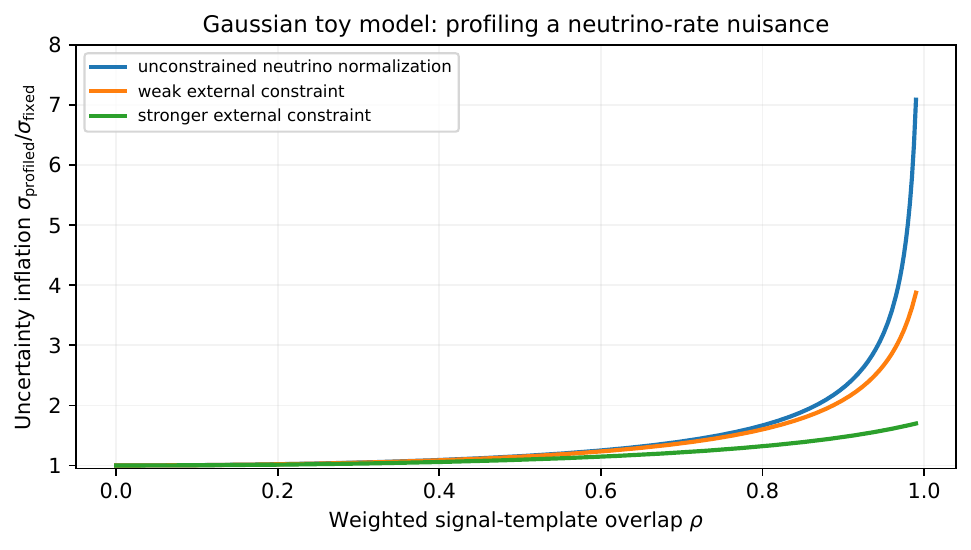}
\caption{Uncertainty inflation from profiling neutrino-background deformations as a function of the weighted spectral overlap $\rho$. Curves correspond to external-information ratios $\lambda=0$, $0.05$, and $0.5$. The calculation uses an idealized Fisher model without experimental detector responses or dark-matter mass assumptions.}
\label{fig:fog}
\end{figure}

Equation~\eqref{eq:degrade} identifies two ways to recover sensitivity: reducing the spectral overlap or improving external constraints on neutrino interactions. Figure~\ref{fig:fog} illustrates this dependence. Nearly identical spectra produce large degeneracies when neutrino parameters are poorly constrained. Stronger external constraints reduce the loss of sensitivity. The calculation is illustrative and does not reproduce experimental discovery limits. Realistic analyses require multiple nuisance parameters, recoil responses, and correlated neutrino-flux uncertainties.

\paragraph*{A V1-derived neutrino deformation.}
The deformation in Eq.~\eqref{eq:fisher} can be tied to a specified interaction.
At fixed mediator mass, the V1 rate of Eq.~\eqref{eq:v1_ratio} gives, for the
single-isotope Xe example,
\begin{equation*}
\begin{split}
\mu_i^\nu(\kappa,m_X)&=\mathcal E\int_i dT\,w(T)[1+\kappa d(T,m_X)]^2,\\
n_i(m_X)&=\left.\frac{\partial\mu_i^\nu}{\partial\kappa}\right|_{0}
=2\mathcal E\int_i dT\,w(T)d(T,m_X),\\
h_i(m_X)&=\mathcal E\int_i dT\,w(T)d^2(T,m_X),\\
\mu_i^\nu(\kappa,m_X)&=b_i+\kappa n_i+\kappa^2h_i.
\end{split}
\tag{17a}\label{eq:physical_template}
\end{equation*}
Here $w(T)$ includes the SM cross section and the solar flux, $b_i=\mu_i^\nu(0)$,
and $\mathcal E$ fixes the normalization. For isotope mixtures the same expressions
hold with the weighted sum of the isotope spectra and their separate $d_a$ inside
the integrals. The linear Fisher model corresponds to $\beta=\kappa$ at the SM
reference point, not to omitting the quadratic term at finite coupling.
The signed weak charge implies $d<0$ and hence $n_i<0$ in all bins of this example.

We use the $^{131}$Xe, $^8$B and Helm inputs of the common V1 calculation on
$0.5$--$3.0\,\mathrm{keV}$, divided into 25 equal bins of width $0.1\,\mathrm{keV}$.
The idealized exposure is chosen so that $\sum_i b_i=N_\nu=1000$ expected SM
events; this is not an observed event count or a detector exposure forecast.
Efficiency is unity, energy reconstruction is exact, the $^8$B spectral shape is
fixed, and other backgrounds are absent. We compare this neutrino deformation
with an isospin-conserving spin-independent \emph{contact-interaction} WIMP of
mass $m_\chi=6\,\mathrm{GeV}$, using
\begin{equation*}
\begin{gathered}
s_i\ \propto\ \int_i dT\,F^2(q)\eta(v_{\min}),\qquad
v_{\min}=\sqrt{\frac{MT}{2\mu_{\chi N}^2}},\qquad
\mu_{\chi N}=\frac{m_\chi M}{m_\chi+M},\\
\eta(v_{\min})=\int_{|\bm v|>v_{\min}}\frac{f_{\rm lab}(\bm v)}{|\bm v|}\,d^3v.
\end{gathered}
\tag{17b}\label{eq:wimp_template}
\end{equation*}
We normalize $\sum_i s_i=1$, so $\alpha$ is the expected WIMP event number.
The Galactic velocity distribution is a Maxwellian proportional to
$\exp(-u^2/v_0^2)\Theta(v_{\rm esc}-u)$, shifted by a constant laboratory speed
$v_E$, with $(v_0,v_{\rm esc},v_E)=(220,544,232)\,\mathrm{km\,s^{-1}}$.
Its normalization is
$N_{\rm esc}=\operatorname{erf}(z)-2z e^{-z^2}/\sqrt\pi$, $z=v_{\rm esc}/v_0$;
the implementation evaluates the corresponding truncated mean inverse speed.
No annual modulation or halo uncertainty is included. This independently
specified WIMP template does not assume that its scattering is mediated by V1:
a light V1 mediator coupled to $\chi$ would introduce an additional recoil
propagator and require recomputing $s_i$.

We calculate $\rho$ with $b_i=\mu_i^\nu(0)$ using bin integrals, and evaluate
Eq.~\eqref{eq:degrade} first with an unconstrained signed $\kappa$, and then
with a hypothetical Gaussian prior $\sigma_\kappa=10^{-11}$ centered on zero.
The same prior is used at every mediator mass; consequently
$\lambda=(\sigma_\kappa^2\sum_i n_i^2/b_i)^{-1}$ is mass dependent.
The results are in Table~\ref{tab:physical_fog}. The negative overlap records
the sign of the interference deformation; it must not be replaced by the overlap
of a positive BSM-amplitude-squared template. Its sign drops out of the
one-nuisance information loss, which depends on $\rho^2$.

\begin{table}[t]
\centering\small
\caption{Physical V1 templates in an idealized Xe experiment. The inflation
$D=\sigma_{\alpha,\rm prof}/\sigma_{\alpha,\rm fixed}$ uses the same completely
fixed-background denominator in every column. $D_{\kappa}$ profiles only
$\kappa$; $D_{\kappa,\zeta}$ also profiles a fractional flux normalization with
$\sigma_\zeta=0.05$. The finite-$\kappa$ prior is $10^{-11}$, whereas the
free-$\kappa$ column has no external prior and fixes the flux.
These are local Fisher uncertainties, not measured confidence intervals or
experimental discovery limits.}
\label{tab:physical_fog}
\begin{tabular}{@{}rrrrrr@{}}
\toprule
$m_X$ [MeV] & $\rho$ & $\lambda$ & $D_\kappa$ (free) & $D_\kappa$ (prior) & $D_{\kappa,\zeta}$ \\
\midrule
1 & -0.94539 & 0.2126 & 3.068 & 1.950 & 2.482 \\
10 & -0.96947 & 0.5315 & 4.078 & 1.609 & 2.236 \\
50 & -0.99684 & 45.1805 & 12.596 & 1.011 & 1.868 \\
\bottomrule
\end{tabular}

\end{table}

For completeness, the last column permits
$\mu_i=b_i+\alpha s_i+\kappa n_i+\zeta b_i$ locally, with independent
Gaussian priors on $\kappa$ and $\zeta$. Writing
$B=\operatorname{diag}(b_i)$, $H=(\bm n,\bm b)$ and
$P=\operatorname{diag}(\sigma_\kappa^{-2},\sigma_\zeta^{-2})$, the Schur complement is
\begin{equation*}
I_{\rm prof}=\bm s^TB^{-1}\bm s
-\bm s^TB^{-1}H(H^TB^{-1}H+P)^{-1}H^TB^{-1}\bm s.
\tag{17c}\label{eq:multi_nuisance_fisher}
\end{equation*}
The product $\zeta\kappa$ in the full flux-rescaled V1 rate is second order and
does not change these derivatives at the reference point.

The heavier-mediator deformation is more similar to the selected WIMP spectrum:
at $1$, $10$ and $50\,\mathrm{MeV}$, $\rho=-0.94539$, $-0.96947$ and $-0.99684$.
With no coupling prior the corresponding inflations are $3.068$, $4.078$ and
$12.596$. A fixed coupling prior reverses that trend, yielding $1.950$, $1.609$
and $1.011$, because the derivative amplitude decreases as the mediator becomes
heavy. In particular, retaining a free coupling at arbitrarily large but finite
mass is not the same statistical limit as removing a decoupled interaction:
without a coupling bound a rate deformation can still be profiled by rescaling
$\kappa$ with $m_X^2$. The finite-prior case correctly returns to unit inflation
when the flux is fixed. With flux uncertainty, a residual loss remains.
These numbers depend on the fixed WIMP mass, window, spectrum, event count and
priors. A discovery analysis additionally needs the full likelihood and the
physical boundary on the nonnegative WIMP strength.

\begin{figure}[!htbp]
\centering
\includegraphics[width=\linewidth]{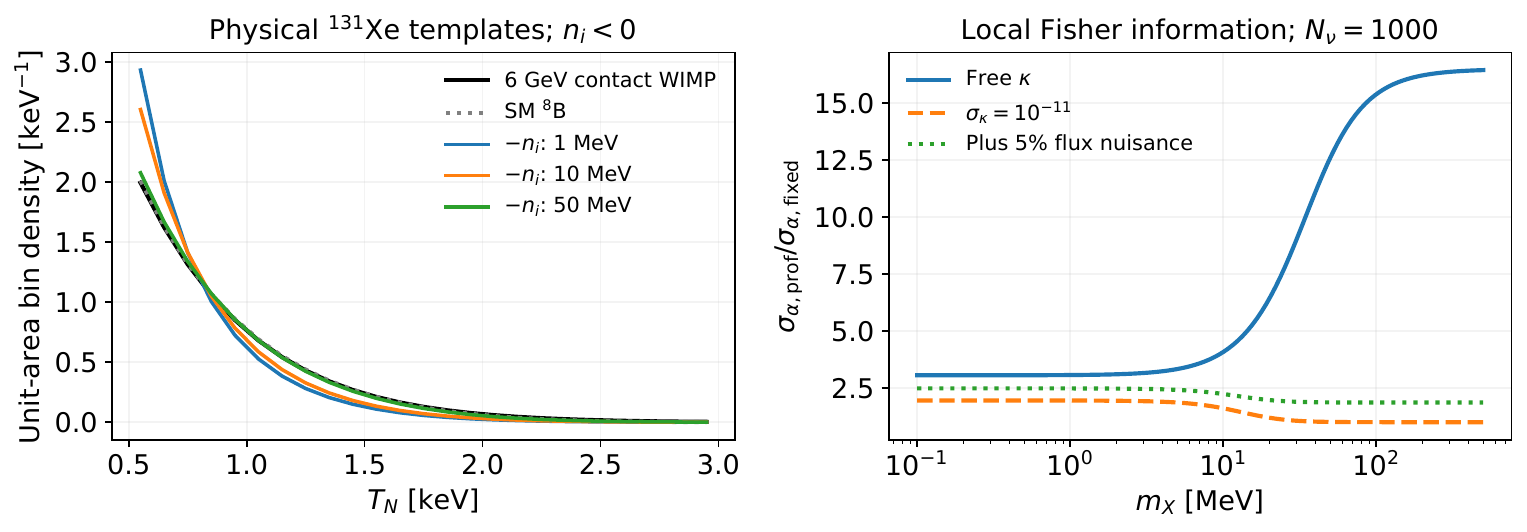}
\caption{V1-derived neutrino deformations and their Fisher impact.
Left: unit-area binned shapes for the fixed $6\,\mathrm{GeV}$ contact WIMP, the
SM solar-$^8$B contribution, and $-n_i$ for the three mediator masses. Plotting
$-n_i$ makes the shapes visible on a positive axis; the calculated overlaps
retain $n_i<0$. Right: uncertainty inflation versus mediator mass for a free
coupling, the fixed coupling prior, and the additional $5\%$ flux nuisance.
The analytical dependence in Figure~\ref{fig:fog} is retained as the reference.
All curves use the idealized inputs stated in the text.}
\label{fig:physical_fog}
\end{figure}

\section{Complementarity Beyond Nuclear Recoils}
\label{sec:other}

Combining different observables can resolve degeneracies in light-mediator couplings. Nuclear recoils constrain products of neutrino and quark couplings, while electron recoils probe $g_\nu g_e$. Neutrino oscillations provide complementary information through forward-scattering matter potentials \cite{Coloma2021,FarzanTortola2018}. Quantum-information studies also discuss decoherence as an alternative modification of neutrino dynamics \cite{Serao2025QINeutrinos}; such effects should not be identified with mediator-induced matter potentials. Different nuclear targets probe distinct proton and neutron charges, while neutrino sources test flavor-dependent interactions \cite{Akimov2021Argon,Ackermann2025,COHERENT2026Ge}. Their combined sensitivity depends on the mediator mass, charge assignments, and relations among couplings.

Precision measurements provide further constraints. Atomic parity violation probes electroweak interactions and nuclear structure \cite{CadedduDordei2019}. However, the purely vector interactions in Eq.~\eqref{eq:vector} do not generate parity-violating electron--quark operators at tree level. Additional axial couplings or electroweak mixing are required. Similarly, constraints on charged fermions can restrict neutrino interactions only when the underlying model relates their couplings. Phenomenological mediators, effective NSI descriptions, and ultraviolet-complete $U(1)'$ models therefore require different interpretations \cite{Coloma2021,Li2026EFT}. Dedicated studies of $X_{17}$ bosons use magnetic moments and atomic spectroscopy to constrain charged-lepton couplings \cite{Serao2025X17Impact,Serao2026g2}.

Displaced $e^+e^-$ vertices from inelastic dark-sector cascades provide a different search channel for light vectors \cite{Serao2026X17Cascade}. Precision interferometry and geometric phases provide further dark-sector probes, with interaction structures distinct from CEvNS \cite{Serao2026ScalarInterferometry,Serao2026GeometricPhases}. DARWIN aims to combine dark-matter searches with precision solar-neutrino measurements \cite{DARWIN2016}. Proposed CEvNS programs at the European Spallation Source and J-PARC could extend the available neutrino sources and nuclear targets \cite{Chattaraj2025ESS,Collar2026JPARC}. The recent XENONnT neutrino--electron measurement also strengthens the prospects for combined nuclear- and electron-recoil analyses \cite{Aprile2026Electron}. These measurements can improve light-mediator constraints by probing independent coupling combinations and reducing parameter degeneracies.

\section{Combining Experimental Constraints}
\label{sec:global}

Light-mediator constraints can be interpreted at three levels. Observable-level analyses constrain event rates using experimental spectra and detector responses. Simplified models specify the mediator interactions, flavor structure, and fermion couplings. Ultraviolet (UV) models impose additional theoretical relations and experimental constraints. Each level requires different assumptions. Model-dependent limits cannot be treated as universal bounds.

A consistent combination requires the likelihood of each experiment and the correlations among shared parameters. The joint likelihood can be written as
\begin{equation}
\label{eq:joint}
\Lagr_{\rm joint}(\bm\theta,\bm\eta)
=
\prod_a
\Lagr_a(D_a\mid\bm\theta,\bm\eta_a,\bm\eta_{\rm shared})
\,\pi(\bm\eta_{\rm shared})
\prod_a\pi_a(\bm\eta_a),
\end{equation}
where $\bm\theta$ contains the mediator parameters, $\bm\eta_a$ denotes experiment-specific nuisance parameters, and $\bm\eta_{\rm shared}$ includes common inputs such as solar-neutrino fluxes and oscillation parameters. The factors $\pi$ represent auxiliary constraints or prior distributions, according to the statistical framework. Shared uncertainties must be included consistently and counted only once. Published confidence intervals cannot be combined directly without the underlying likelihood information.

The cases V1, V2, and S1 can be used to organize experimental reinterpretations. For the vector cases, the interference terms, neutrino flavor composition, and relative coupling signs must be specified. For the scalar case, the required inputs are hadronic matrix elements, recoil-dependent cross sections, and final-state neutrino properties. In all cases, the predicted rates must include source spectra, nuclear or atomic responses, and detector effects. Table~\ref{tab:recast} summarizes the required inputs.

\begin{table}[t]
\centering
\small
\caption{Inputs required to reinterpret light-mediator constraints for cases V1, V2, and S1.}
\label{tab:recast}
\begin{tabularx}{\textwidth}{@{}p{.16\textwidth}p{.24\textwidth}X@{}}
\toprule
Stage & Application & Required information \\
\midrule
Interaction &
V1/V2/S1 &
Mediator couplings, flavor structure, interference terms, and propagator dependence \\
Source &
Stopped-pion, reactor, solar &
Neutrino spectra, timing, flavor composition, and flux uncertainties \\
Target &
Ge/Ar/CsI/Xe &
Isotopic composition, nuclear form factors, scalar charges, and associated uncertainties \\
Observable &
Nuclear/electron recoils &
Quenching, ionization and scintillation yields, resolution, and detection efficiency \\
Likelihood &
All datasets &
Event distributions, backgrounds, nuisance parameters, statistical prescriptions, and correlations \\
Validation &
All cases &
Reproduction of published SM predictions and published constraints for the chosen model \\
\bottomrule
\end{tabularx}
\end{table}

Reliable global constraints require validated event-rate predictions and consistent statistical inference. Public experimental data, including the COHERENT release \cite{COHERENTRelease}, provide inputs for testing these calculations. The three interaction models and the accompanying scripts make the assumed coupling dependencies and degeneracies explicit. Combined exclusion limits require further validation against experimental likelihoods.

\section{Conclusions}
\label{sec:conclusions}

Light-mediator searches now connect neutrino scattering, precision measurements, and dark-matter direct detection. However, experimental constraints are complementary only when they probe independent combinations of mediator parameters. Their interpretation requires consistent interaction models, nuclear responses, and statistical assumptions.

We have compared scalar and vector interactions in the momentum-transfer ranges relevant to neutrino and dark-matter detectors. Their recoil spectra depend on mediator mass, Lorentz structure, and momentum transfer. Vector interactions can interfere with the Standard Model, while chirality-changing scalar interactions produce distinct recoil signatures. Nuclear and electron recoils constrain different coupling products unless additional model relations are imposed. Different nuclear targets do not necessarily resolve coupling degeneracies. For representative Ge and Xe isotopes, the nuclear-charge response matrix has a condition number of approximately 115, revealing nearly parallel blind directions.

The weighted charge analysis separates nuclear geometry from assumed precision and shows that a Ge--Ar pair can be more degenerate than Ge--Xe. The common V1 calculation quantifies source-dependent recoil and integrated-rate changes while keeping the coupling product fixed; these changes are not experimental sensitivity rankings.

The neutrino-fog calculation quantifies the effect of uncertain neutrino interactions on dark-matter discovery sensitivity. The local Fisher model identifies spectral overlap and neutrino-background uncertainties as the main sources of sensitivity loss. A V1-derived solar-neutrino template makes this dependence explicit for a fixed WIMP mass and shows why a fixed coupling prior and an unconstrained coupling produce different mediator-mass trends. Measurements that constrain the neutrino contribution independently can reduce this loss of sensitivity; the neutrino fog is not an absolute discovery limit.

A combined analysis of light-mediator searches must specify the interaction model and its experimental inputs. Combining experiments requires consistent coupling definitions, momentum-dependent cross sections, detector responses, and correlated uncertainties. Adding datasets is useful only insofar as they constrain independent directions in parameter space. This criterion can guide the choice of observables and targets for future neutrino and dark-matter measurements.

\vspace{6pt}

\authorcontributions{The authors contributed equally to the conceptualization, analysis, writing, and revision of the manuscript.}

\acknowledgments{The authors acknowledge the support of Istituto Nazionale di Fisica Nucleare (INFN), Iniziative Specifiche QGSKY.}
\conflictsofinterest{The authors declare no conflicts of interest.}

\ifdefined\MDPIFallback
  \bibliographystyle{unsrtnat}
  \bibliography{references}
\else
  \begin{adjustwidth}{-\extralength}{0cm}
  \bibliography{references}
  \PublishersNote{}
  \end{adjustwidth}
\fi
\end{document}